\documentclass[twocolumn,aps,prb,superscriptaddress]{revtex4-2}

\usepackage{graphicx}
\usepackage{color}
\usepackage{amsmath}
\usepackage{upgreek}

\usepackage{todonotes}

\newcommand{\cdag}{c^{\dagger}}%

\begin{document}

\title{Computation Kernel for Feynman Diagrams}
\author{Daria Gazizova}
\affiliation{Department of Physics and Physical Oceanography, Memorial University of Newfoundland, St. John's, Newfoundland \& Labrador, Canada A1B 3X7} 
\author{Rayan Farid}
\affiliation{Department of Physics and Physical Oceanography, Memorial University of Newfoundland, St. John's, Newfoundland \& Labrador, Canada A1B 3X7} 
\author{B. D. E. McNiven}
\affiliation{Faculty of Engineering and Applied Science, Memorial University of Newfoundland and Labrador, St. John's, Newfoundland \& Labrador, Canada A1B 3X5} 
\affiliation{Compute Everything Technologies Ltd., St. John's, Newfoundland \& Labrador, Canada}
\author{I. Assi}
\affiliation{Department of Physics and Physical Oceanography, Memorial University of Newfoundland, St. John's, Newfoundland \& Labrador, Canada A1B 3X7} 
\author{Ethan G. Armstrong}
\affiliation{Compute Everything Technologies Ltd., St. John's, Newfoundland \& Labrador, Canada}
\author{J. P. F. LeBlanc}
\email{jleblanc@mun.ca}
\affiliation{Department of Physics and Physical Oceanography, Memorial University of Newfoundland, St. John's, Newfoundland \& Labrador, Canada A1B 3X7} 
\affiliation{Compute Everything Technologies Ltd., St. John's, Newfoundland \& Labrador, Canada}

\date{\today}
\begin{abstract}
We present a general representation for solving problems in many-body perturbation theory.  By projecting the single-particle Green's function to an auxiliary space we show how one can convert an arbitrary Feynman graph to a universal kernel representation.  Once constructed, the computation kernel contains no problem specific information yet contains all explicit temperature and frequency dependence of the diagram.  This computation kernel is problem agnostic, and valid for any physical problem that would normally leverage the Matsubara formalism of many-body perturbation theory. 
The result of any diagram can be written as a linear combination of these computation kernel elements with coefficients given by a sum over products of known tensor elements that are themselves problem specific and represent spatial degrees of freedom.  We probe the efficacy of this approach by generating the computation kernel for a low order self-energy diagram which we then use to construct solutions to distinct problems.
\end{abstract}

\maketitle


\section{Introduction}
The need for precise evaluation of correlated electron problems drives the development of numerous numerical approaches and serves as a catalyst for advancing quantum computational hardware and algorithms.  
Despite advancements in algorithms for quantum hardware, the need for purely classical approaches remains undiminished, as they have accounted for virtually all useful computations of correlated electron systems to date.\cite{carleo:2024,waintal:2024}
On classical hardware, quantum algorithm development spans several research areas, including the study of equilibrium and non-equilibrium systems, methods for finite temperature and ground state properties, and approaches using wavefunctions versus correlation functions. All these avenues are worth pursuing, with the shared goal of solving challenging quantum systems on classical or quantum hardware.\cite{mingpu:2022} 

Among the methods for finite temperature systems in equilibrium is many-body perturbation theory that is formulated in terms of Feynman diagrams.  
While not a new concept, Feynman diagrams remain a cornerstone of computational physics and chemistry. In quantum chemistry, the lowest-order perturbations, Hartree-Fock theory, serve as the foundation for many approaches and can be formulated in terms of diagrams.  Going beyond Hartree-Fock, diagrammatic methods are valuable not only due to their broad applicability but also for their potential to provide a path to finite-temperature results for dynamic properties while offering the flexibility to evaluate a variety of observables.  These methods are, however, at least exponential in complexity as the order of the interaction increases,\cite{Rossi,RossiCDet2017,Evgeny:2019} and self-consistent approaches have shown evidence of false convergence\cite{kozik:2015LW,kozik:2025,mcniven:2021}.

At the same time there have been algorithmic advancements intended to drastically improve computation time.  In the case of real and imaginary time representations, application of tensor methods (tensor trains, tensor-cross interpolation) have shown promise for reducing the complexity of numerical problems by reducing expressions to tensor products\cite{ttrains1,erpenbeck:ttrain,tci,ttrain:2particle, hiroshi:2025prl}.  Other alternate representations for Matsubara Green's functions show similar promise for application to diagrammatic problems\cite{markus2,markus:ir, kaye:dlr, kaye:3point, kaye:prx:2024, shinaoka:jpsj}.  
In the case of problems formulated in the Matsubara frequency domain there are also symbolic approaches that have emerged and are able to generate analytic solutions to internal temporal degrees of freedom of arbitrary Feynman diagrams\cite{jaksa,jaksa:imag,jaksa:conductivity,AMI}.  One such symbolic approach, algorithmic Matsubara integration (AMI), has been applied to a broad spectrum of problems\cite{AMI,taheri:2020,GIT,Assi,mcniven:2021,mcniven:2022,farid:2023,farid:2024,leblanc:2022,Gazizova(2023),burke:2023}.
However, the AMI approach is currently limited in its application for a number of reasons.  Computationally, the method suffers from a nearly factorial growth of the number of function calls required to evaluate diagrams of increasing order limiting applications to fifth or sixth order in practice\cite{leblanc:2022,mcniven:2021}.  The method also requires access to the pole structure of the Green's function which prevents one from merging AMI with self-consistent schemes or applying AMI to systems with non-diagonal Green's functions.  

Alternate representations of Green's functions provide a possible path forward for the AMI approach.  In particular, the discrete Lehmann representation (DLR) is an alternate function representation for Matsubara Green's functions in both the imaginary time and imaginary frequency domains.  The DLR represents the Green's function in imaginary frequency via a discretized spectral kernel wherein the pole structure is symbolically accessible and this allows one to merge the DLR with AMI to provide self-consistent schemes.  This was recently shown to be effective for studying single-band impurity problems in both single-shot and self-consistent schemes.\cite{dgaz:2024} There it was shown that an exponential trade-off in computational expense is accompanied by an exponential convergence of the result to the numerically exact result.  However, applying this approach to multi-band or finite systems scales poorly due to the addition of yet another exponential computational scaling on top of existing factorial scalings.  

In this work, we make substantial progress in applying the DLR combined with AMI to arbitrary systems beyond impurities such as multi-band problems or lattice problems.  We consider the implications of applying the DLR to Feynman diagrams with additional internal spatial degrees of freedom and find that there is a factorizable component of the problem that is identical to the case of an impurity problem but depends only on the non-physical space of DLR poles.  This realization allows us to rephrase the factorizable component as a kernel for systems with any set of internal spatial degrees of freedom.  We call this factorizable component the computation kernel.  Once the computation kernel is computed for a given diagram, it can be reused endlessly for new problems regardless of the spatial geometry of the system.  We show that diagrams can be decomposed as a linear combination of computation kernel entries with coefficients that are a product of tensor elements that are \emph{apriori} known from the DLR fit of the single-particle Green's function. 

\section{General Methodology}
\subsection{Discrete Lehmann Representation}
This work relies heavily on the formulation of the discrete Lehmann representation (DLR) presented in Refs.~\cite{kaye:dlr,kaye:libdlr} for single particle Green's functions. 
The premise of the DLR is to approximate the spectral decomposition in the truncated form,
\begin{equation}\label{eq:spec}
    G(i\omega_n)=\int\limits_{-\Lambda}^{\Lambda} K(i\omega_n,x)\rho(x) dx,
\end{equation}
where $\rho$ is the density and the kernel is given by $K(i\omega_n,x)=\frac{1}{i\omega_n-x}$ when represented in Matsubara frequencies, $i\omega_n$.
In the frequency domain, the DLR procedure assigns a set of nodes along the imaginary axis, $\{i\omega_{DLR}\}$, that are determined uniquely for a given choice of $\Lambda$ for a desired error tolerance at a chosen inverse temperature, $\beta$.  The coefficients are selected such that the kernel is approximated with an error tolerance, $\upvarepsilon$, in the domain $[-\Lambda,\Lambda]$ based on fitting the discrete set of poles along the real axis, $\{x_\ell \}$.
 The Matsubara Green's function can then be approximated as a sum over a finite set of poles $\{x_\ell \}$ with weights $A^\ell$ as
\begin{equation}\label{eq:gdlr}
    G(i\omega_n) \approx G_{DLR}(i\omega_n)=\sum\limits_{\ell=1}^{r} K(i\omega_n, x_\ell) A^\ell.
\end{equation}
We pause to discuss an important biproduct of Eq.~(\ref{eq:gdlr}).  Take for example a simple scenario of a non-interacting Green's function of a single pole, $G(i\omega_n,\epsilon)=\frac{1}{i\omega_n-\epsilon}$.  Once we evaluate  $G(i\omega_n,\epsilon)$ along the set of $\{i\omega_{DLR}\}$ frequencies, the DLR fit provides an approximation via the right-hand-side of Eq.~(\ref{eq:gdlr}) that no longer explicitly contains $\epsilon$.  Computationally, one is worse off in this scenario because of the replacement of a single physical simple pole, $\epsilon$, with an entire set of auxilliary poles $\{x_\ell\}$.  However, mathematically, the removal of $\epsilon$ from the problem provides a potential enormous advantage, allowing us to write very simple, yet general expressions for the evaluation of Feynman diagrams for arbitrary problems.

Once the set of $A^\ell$ coefficients are obtained, the Green's function can be used without knowledge of the physical pole that gave rise to the Green's function. One can view this as the replacement of a physical space of poles with a computational space that is problem agnostic and therefore identical for all problems.  

\subsection{Algorithmic Matsubara Integration}
Algorithmic Matsubara Integration is a symbolic tool to generate the closed-form analytic solution to sequences of Matsubara sums for products of Green's functions.  This procedure is identical to the method one would follow by hand, and relies on a single residue identity for Fermionic frequency summation
\begin{equation}
    \frac{1}{\beta}\sum\limits_{i\nu_n}H(i\nu_n) = \sum\limits_{\{z_0\}} \text{Res}[f(z) H(z)]_{z_0},
\end{equation}
where $H(i\nu_n)$ and $\{z_0\}$ are the poles of the function $H$ with respect to $i\nu_n$ and $f(z)\equiv f(z,\beta)=1/(1+e^{\beta z})$ is the Fermi function.  If the poles of the function can be uniquely determined then the Matsubara summation can be trivially performed analytically.  The result of such an analytic summation is well-known to be in the form of the original problem, meaning that a sequence of sums over distinct Matsubara summations can be performed analytically, resulting in a total number of terms equal to the product of the number of residues at each integration step. Performing AMI requires a more general symbolic representation of the functions to allow Green's functions to carry arbitrary linear combinations of the frequency and energy arguments.  Green's functions then shift from being locally defined to being globally aware of all frequencies arising in a diagram. Although not widely adopted, the approach has been applied to diagrammatic expansions for self-energies and susceptibilities up to 6th order which requires the evaluation of between $10^5\to 10^6$ analytic expressions.\cite{mcniven:2021, farid:2023}

Sadly, for many numerical diagrammatic approaches, each Green's function comprising $H$ is not known analytically but is instead known at only the Matsubara points and this prevents the application of AMI.  The discrete Lehman representation solves this issue for the AMI methodology by providing a basis of functions with known pole structure.  Recently, it was shown that combining the two methods remains numerically stable.\cite{dgaz:2024}   

\subsection{Extending the DLR for use with AMI}

\begin{figure}
    \centering
    \includegraphics[width=0.6\linewidth]{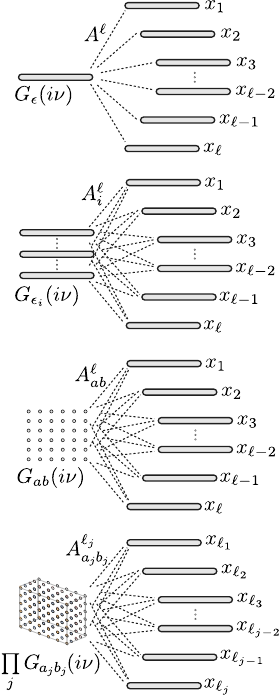}
    \caption{Schematic representation of applications of a discrete Lehmann representation to increasingly complex datasets.  From top to bottom these represent: an impurity Green's function, a diagonal Green's function, a matrix-valued Green's function, and a collection of Green's functions (as appear in Feynman diagrams).  On the left of each schematic is a physical space of states while on the right is a particular basis of DLR poles, $x_\ell$.   }
    \label{fig:schematic}
\end{figure}

Before deriving general expressions, we take a pedantic approach and walk through precisely what the DLR representation is doing for a sequence of problems of increasing complexity.  To aid in this discussion, we turn the reader's attention to the schematic of Fig.~\ref{fig:schematic}.  What we are depicting there is a sequence of problems of increasing complexity moving from top to bottom.  The first scenario is the decomposition of a Matsubara Green's function  of a physical state, $\epsilon$, (depicted as a bar on the left) into a DLR representation\cite{kaye:libdlr,kaye:cppdlr} of new auxilliary states $x_\ell$ (depicted as a set of bars on the right).  This is what is described in Eq.~(\ref{eq:gdlr}), where a physical Green's function is now a sum over a non-physical space of non-interacting Green's functions with a set of coefficients, $A^\ell$, as weights. Importantly, the non-physical states on the right, are no-longer explicitly dependent on the physical state $\epsilon$. 

If one can find an appropriate set of weights for the first problem, equivalent to a quantum impurity, then increasing the complexity of the physical space is straightforward.  The second panel of Fig.~\ref{fig:schematic} shows a set of physical eigenstates, $\epsilon_i$, each of which may have its own frequency dependent Green's function.  Using the DLR approach, one can sequentially project each physical Green's function into a fixed DLR basis.  All that is required in doing so is tracking an additional index on the weight coefficients, $A^\ell_i$ for each of the $\epsilon_i$ eigenstates.  Again, the non-physical states on the right are no-longer explicitly dependent on the physical eigenstates $\epsilon_i$.  Further, since we have chosen to project all of the physical states into a single fixed DLR basis, the mathematical representation of any choice of physical problem will now be mathematically identical and only the weight values will change.

Continuing the argument is straightforward for more complex problems.  In the case of matrix-valued single particle Green's function, $G_{ab}$, the usual DLR representation is applied and all matrix elements are assigned coefficients for the same set of $x_\ell$ poles.  The storage of the necessary information requires a coefficient tensor that matches the geometry of the matrix representation of the Green's function of the form $A^\ell_{ab}$ for Green's function elements $G_{ab}(i\omega_n)=\langle c_a(\tau)c^\dagger_b(0) \rangle$.  

For the present work, we need to simultaneously store all information for Feynman diagrams that are at their core just collections of matrix-valued Green's functions as depicted in the bottom panel of Fig.~\ref{fig:schematic}.  In this case, one can represent a collection of Green's functions, each with an index $j$, as $G^{(j)}$ either all in the same DLR representation or as a collection of unique representations by increasing the rank of the coefficient tensor.  We generate a set of real-frequency DLR nodes for each of the $j$ Green's functions, $\{x_{\ell_j}\}$, and then store their coefficients as $A^{\ell_j}_{a_jb_j}$. One can then represent a particular element of the $j$th Green's function in the DLR representation in the form,
\begin{equation}\label{eq:gdlr_matrix}
    G_{a_jb_j}(i\omega_n)=\sum\limits_{\ell_j}\frac{A^{\ell_j}_{a_jb_j}}{i\omega_n - x_{\ell_j}}.
\end{equation}
In the case that all DLR frequency nodes are the same this could be represented for all Green's functions via a single rank-4 tensor, $A_{j\ell ab}$.  However, the AMI method relies on residue theory and the use of a fixed DLR representation results in multi-poles.  While this can be treated via multi-pole residue theorem it changes the analytic expression for the kernel.  This can be avoided by taking the form of Eq.~(\ref{eq:gdlr_matrix}) and assume the entries of $\{x_{\ell_j}\}$ are all distinct and hence multipoles are avoided.   We note that our procedure is not a particularly novel use of the DLR approach, nor is it intended to be, and emphasize that it requires only repeated applications of existing DLR algorithms for matrix-value Green's functions.\cite{kaye:libdlr,kaye:cppdlr}

\subsection{Application to Feynman Diagrams}
Having now established the notation for the replacement of Green's functions in a physical space with those in an arbitrary DLR representation we now move to the problem at hand of applying this to Feynman diagrams.
Prior to applying the DLR form of the Green's function, a Feynman diagram is represented in a physical space of energy levels and is determined by a product Green's functions.  Depending on the observable the diagram represents, there will typically be a number, $m$, independent labels, and, $n$, dependent labels.  
If one multiplies out the matrix-valued Green's functions, then a diagram is a high-dimensional summation over products of elements of those Green's functions.  Focusing on a single term of such products, we define 
\begin{widetext}
\begin{equation}
    H(\{i\nu_m\}, i\nu_x, \{a_j,b_j\})=G_{a_1b_1}(i\nu_1)... G_{a_mb_m}(i\nu_m)G_{a_{m+1}b_{m+1}}(\vec{\alpha}_{1}\cdot\vec{\nu})...G_{a_{m+n}b_{m+n}}(\vec{\alpha}_{n}\cdot\vec{\nu}),  
\end{equation}
\end{widetext}
where $\{i\nu_m\}$ is the set of internal Matsubara frequencies, $i\nu_x$, is one or more external Matsubara frequency, and the vector $\vec{\nu}=\{i\nu_1,...,i\nu_m,i\nu_x\}$ is defined such that the dot products  $\vec{\alpha}_i\cdot\vec{\nu}$ represent an arbitrary linear combination of internal and external frequencies that conserve energy and depend on diagram topology.\cite{AMI}
With these simplifications, a particular diagram is then a summation of the independent Matsubara frequencies $i\nu_1\to i\nu_m$ as well as a summation over the relevant set of of matrix indices $\{a_j,b_j\}$ which depends on diagram topology.  One must also multiply each term in that sum by an appropriate product of interaction matrix elements which we denote for now as $\bar{U}$.   
In the case of quantum impurity problems, $\bar{U}$ will be a product of interacting matrix elements, while for lattice problems it might contain products of interacting potentials $\prod V(\mathbf{q})$. 
Regardless, we treat the detailed interpretation of the summation over indices $\{a_j,b_j\}$  and detailed function dependence of $\bar{U}$ as problem specific, and revisit in detail later. 

A particular diagram topology $\mathcal{D}$ will then contain terms in the form
\begin{equation}
    \mathcal{D}=\sum\limits_{\{i\nu_m\}}\sum\limits_{\{a_j,b_j\}} H(\{i\nu_m\}, i\nu_x, \{a_j,b_j\})\bar{U}, 
    \label{eq:primary1}
\end{equation}
where we have, for now, suppressed the dependence of $\mathcal{D}$ on external parameters.
Our goal is to evaluate Eq.~(\ref{eq:primary1}) and leverage the DLR representation.  By simply replacing every Green's function with its own DLR representation from Eq.~(\ref{eq:gdlr_matrix}), we can rewrite Eq.~(\ref{eq:primary1}) as
\begin{equation}
    \mathcal{D}=\sum\limits_{\{\ell\}}\sum\limits_{\{i\nu_m\}}\sum\limits_{\{a_j,b_j\}} 
     \prod\limits_{i=1}^{m} \frac{A^{\ell_i}_{a_i b_i} }{i\nu_i-x_{\ell_i}}
     \prod\limits_{j=1}^{n} \frac{A^{\ell_{j+m}}_{a_{j+m}b_{j+m}}}{\vec{\alpha_j}\cdot \vec{\nu}-x_{\ell_{j+m}}}
    \bar{U}, 
    \label{eq:primary2}
\end{equation}
where $\{\ell\}=\{\ell_1,...,\ell_{n+m}\}$.  
We notice that there is a separable component that no longer contains problem specific information, but only the kernels of the DLR representations.  We can therefore rewrite the expression as
\begin{widetext}
\begin{equation}
    \mathcal{D}=\sum\limits_{\{\ell\}}\Bigg[\underbrace{\sum\limits_{\{i\nu_m\}} 
     \prod\limits_{i=1}^{m} \frac{1}{i\nu_i-x_{\ell_i}}
     \prod\limits_{j=1}^{n} \frac{1}{\vec{\alpha_j}\cdot \vec{\nu}-x_{\ell_{j+m}}}}_{\rm Exactly \ Solvable} \sum\limits_{\{a_j,b_j\}} \prod\limits_{k=1}^{m+n} A^{\ell_k}_{a_k b_k} \bar{U}\Bigg]. 
    \label{eq:primary3}
\end{equation}
\end{widetext}
We have underlined the exactly solvable part in Eq.~(\ref{eq:primary3}).  Specifically the Matsubara summation can be performed by hand for simple diagrams or automatically via algorithmic Matsubara integration (AMI).\cite{libami,torchami}  We proceed using AMI and will annotate for a specific diagram this quantity as
\begin{equation}
   \mathcal{K}^{\mathcal{D}}(i\nu_x,\{\ell\},\beta)= \sum\limits_{\{i\nu_m\}} 
     \prod\limits_{i=1}^{m} \frac{1}{i\nu_i-x_{\ell_i}}
     \prod\limits_{j=1}^{n} \frac{1}{\vec{\alpha_j}\cdot \vec{\nu}-x_{\ell_{j+m}}}.
\end{equation}
We emphasize the notational change to mark  that $ \mathcal{K}^{\mathcal{D}}$ is an explicit function of inverse temperature $\beta$ after Matsubara summation.  By substituting into Eq.~(\ref{eq:primary3}) we obtain 
\begin{equation}
    \mathcal{D}(i\nu_x,\beta)=\sum\limits_{\{\ell\}}\mathcal{K}^{\mathcal{D}}(i\nu_x,\{\ell\},\beta) \sum\limits_{\{a_j,b_j\}} \prod\limits_{k=1}^{m+n} A^{\ell_k}_{a_k b_k} \bar{U}. 
    \label{eq:primary_pre}
\end{equation}
We can then replace the summation on the right with an appropriately indexed constant.  We shift external parameters to indices and now specify also the diagram's dependence on those external properties on the left-hand side, which yields
\begin{equation}
    \mathcal{D}(i\nu_x,\beta)=\sum\limits_{\{\ell\}}\mathcal{K}^{\mathcal{D}}_{i\nu_x,\beta}(\{\ell\})\hspace{2pt}  C^{\mathcal{D}}_{\beta}(\{\ell\}). 
    \label{eq:primary_final}
\end{equation}
Equation~(\ref{eq:primary_final}) is a primary result of this paper.  One can interpret the space of $\{\ell\}$ as a kernel for computing the diagram where $\mathcal{K}^{\mathcal{D}}_{i\nu_x,\beta}(\{\ell\})$ represents the kernel functions containing all frequency dependence while containing no problem specific information.  The coefficients $C^{\mathcal{D}}_{\beta}(\{\ell\})$ contain all problem specific information, and while these coefficients depend implicitly on the inverse temperature, $\beta$, they do not depend on frequency and only depend explicitly on the $A^{\ell_k}_{a_k b_k}$ coefficients and the interaction matrix elements. Both the kernel entries, $\mathcal{K}^{\mathcal{D}}$, and the coefficients, $C^{\mathcal{D}}$, depend on a specific diagram topology.  The final summation over the set of $\{\ell\}$ can be linearized such that the result of the diagram is a simple dot-product.  

In summary, by using the DLR representation in a fixed basis, we are able to extract a factorizable component of the problem that only depends on the choice of the DLR basis which we refer to as the computation kernel.  The computation kernel can be evaluated for any particular diagram via algorithmic Matsubara integration in a symbolic form, which can then be evaluated for a choice of external frequencies and inverse temperatures where each result is a single complex-valued number.  If one stores those results as a function of the sets of DLR pole indices, $\{\ell\}$, then, via Eq.~(\ref{eq:primary_final}), one only needs to generate the coefficients to evaluate the diagram.  Since all problem specific information is contained in the coefficients, the kernel is reusable for distinct physical problems.  The computational effort to compute any diagram for any physical problem is then only the evaluation of the non-temporal degrees of freedom which appear as products of the DLR tensor and interaction matrix elements.   
We highlight that, while distinct, the present work has connections with a number of important research advancements including: overcomplete representations, partial spectral functions, and tensor trains.\cite{kernel:halbinger, kernel:kugler,wallerberger:overcomplete, ttrains1, ttrain:2particle}

\subsection{Computational Complexity Analysis}
We analyze the computational complexity of our method by separating the workflow into three phases: (1) the temporal kernel construction via symbolic Matsubara integration, (2) the evaluation of problem-specific coefficients, and (3) final contraction.
For this purpose we define the following:
\begin{itemize}
    \item $m$ = diagram order (number of interaction lines),
    \item $r$ = number of DLR poles per Green’s function (assumed fixed),
    \item $n_G $ = number of Green’s functions in the diagram (assume self energy where $n_G= 2m - 1$),
    \item $L$ = number of orbitals (molecular) or lattice points per dimension (lattice),
    \item $d$ = spatial dimension (for lattice problems).
\end{itemize}
\subsubsection{Temporal Kernel Precomputation (One-Time Cost)}

The kernel $K_D(i\nu_x, \{\ell\}, \beta)$ is evaluated using algorithmic Matsubara integration (AMI), which performs symbolic contour integration over Matsubara frequencies. Each internal frequency results in a residue expansion that increases the number of terms in the expression and this consideration is topology dependent.  However, if one restricts discussion to self-energy diagrams, and generates an empirical average evaluation time for random inputs averaged over all diagrams of order $m$, this leads to a growth in cost roughly proportional to $\mathcal{O}(c^{m})$, with $c \equiv 10^{0.8}\approx 6.3$ based on empirical benchmarks from our implementation\cite{torchami} for diagrams ranging from order 2$\to 9$.

Each kernel entry corresponds to a combination of $r^{n_G} = r^{2m - 1}$ DLR indices, leading to a total kernel evaluation cost:

$$
\mathcal{O}(r^{2m - 1} \cdot 6.3^{m}).
$$

This is done once per diagram for each choice of $\beta$ and can be reused for all subsequent calculations of that diagram for new problems or iterative schemes.

\subsubsection{ Coefficient Evaluation (Per-System Cost)}

After kernel generation, diagram evaluation for a given system requires computing the coefficients $C^{(D)}_\beta(\{\ell\})$, which are tensor contractions over system-specific DLR weights and interaction matrix elements.

\textbf{Molecular problems:}
  The number of terms in the summation over spatial indices scales as $\mathcal{O}(L^{4m - 2})$, and each term involves a product of $2m - 1$ DLR coefficients. Total cost:

  $$
  \mathcal{O}(L^{4m - 2} \cdot r^{2m - 1}).
  $$

\textbf{Lattice problems:}
  For a system of size $L^d$, the spatial summation grows as $\mathcal{O}(L^{dm})$. Total cost:

  $$
  \mathcal{O}(L^{dm} \cdot r^{2m - 1}).
  $$

These contractions are parallelizable and dominate the cost in large systems.

\subsubsection{Final Kernel-Contraction (Negligible Cost)}

Once the kernel and coefficients are prepared, the final diagram value is a dot product over the DLR index space:

$$
\mathcal{O}(r^{2m - 1}).
$$

This step is lightweight and does not contribute significantly to runtime.

\subsection{Comparison to Other Techniques}
Evaluating Feynman diagrams is a central task in finite-temperature many-body calculations. It requires the internal summation over spatial and temporal degrees of freedom.  The most naive numerical summation discretizes each internal degree of freedom.  If spatial degrees of freedom are represented by non-diagonal matrices of geometry $L$ and temporal integrals require $N$ samples for a target accuracy then the naive scaling of the problem is exponential as $\mathcal{O}(L^{4m-2}N^m)$ for a diagram of order $m$ with no opportunity to disentangle the degrees of freedom.  The result of such a calculation might provide only an estimate of the temporal integration resulting in an intrinsic error.  One can eliminate that error with our approach at a total cost of $\mathcal{O}( [L^{4m-2}+c^m]\cdot r^{2m-1})$.

This is of course an active area of research.  To avoid slow convergence of Matsubara frequency sums  many modern approaches turn to the imaginary-time domain. In particular, the intermediate representation (IR)\cite{shinaoka:2022} offers an optimally compact set of basis functions to represent Green’s functions and response functions. By projecting to IR, one can reduce the number of significant time or frequency components to a small number, often as few as $20$–$40$, with controllable accuracy. This significantly lowers the effective cost of Matsubara summations and has enabled efficient impurity and ab initio solvers.  The discrete Lehmann representation we employ has been used elsewhere\cite{kaye:prx:2024, kaye:arxiv} combined with tensor network methods and sum of exponentials fitting procedures leading to drastic reduction in computational complexity.  

If one were to combine the present work with tensor network methods, in particular SVD and tensor-train decompositions, we expect that the primary cost will be the contraction of the network and that the scaling of the approach can be reduced drastically if low-rank structure of the tensors exists\cite{hiroshi:2025prl}.  Nevertheless, the separation of frequency and spatial structures combined with reuse potential of the kernel provides conceptual and practical gains that we will illustrate. 

\section{Problem Specific Application}
We demonstrate a minimal application of the computation kernel approach of Eq.~(\ref{eq:primary_final}). We solve two model Hamiltonians: a two-band impurity with a general 4-index interaction tensor, and a lattice problem, the Hubbard Hamiltonian on the 2D square lattice.  We will restrict ourselves to the evaluation of a single second-order self energy diagram.  We emphasize that the goal is not to extract physically relevant information about these model systems but to demonstrate that the construction of a single computational kernel, $\mathcal{K}^{\mathcal{D}}_{i\nu_x,\beta}(\{\ell\})$, will allow us to evaluate the full result of these two very different physical systems.  

\subsection{Hamiltonians}
\subsubsection{Multi-band Impurities}
We will evaluate diagrams for a very general two-body Hamiltonian
with two terms; a single-particle term, $H_0$, and a generalized four-operator interaction term, $H_V$. These are given by 
\begin{equation}
H =
\underbrace{\sum_{ab} h_{ab}\cdag_a c_b}_{H_0} +
\underbrace{\frac{1}{2} \sum_{abcd} U_{abcd} \cdag_a \cdag_c c_d c_b}_{H_V}.\vspace{10pt}
    \label{eq:h}
\end{equation}
Here $a$ and $b$ are arbitrary band indices and the $c_i^\dagger$ and $c_i$ represent standard creation and annihilation operators in the state $i$, respectively, and the values of $h_{ab}$ represent one-electron integrals while $U_{abcd}$ is the two-electron interaction matrix.  We will use the H$_2$ atom at equilibrium separation in the sto-6G basis as a test case.  We obtain the single-particle matrix $h_{ab}$ based on the restricted Hartree-Fock approximation and the two-electron coulomb integrals $U_{abcd}$ both in the molecular orbital basis in units of hartrees using the \texttt{pyscf} package\cite{pyscf}. Values used for $h_{ab}$ and $U_{abcd}$ are included in the supplementary files \cite{supp}. 

\subsubsection{Hubbard Hamiltonian on a Lattice}
We study the single-band Hubbard Hamiltonian on a 2D square lattice\cite{benchmarks},
\begin{eqnarray}\label{E:Hubbard}
H = \sum_{ ij \sigma} t_{ij}c_{i\sigma}^\dagger c_{j\sigma} + U\sum_{i} n_{i\uparrow} n_{i\downarrow},
\end{eqnarray}
where $t_{ij}$ is the hopping amplitude, $c_{i\sigma}^{(\dagger)}$ ($c_{i\sigma}$) is the creation (annihilation) operator at site $i$, $\sigma \in \{\uparrow,\downarrow\}$ is the spin, $U$ is the onsite Hubbard interaction, and $n_{i\sigma} = c_{i\sigma}^{\dagger}c_{i\sigma}$ is the number operator.  We restrict the sum over sites to nearest neighbors for a 2D square lattice, resulting in the free particle energy 
\begin{eqnarray}
\nonumber\epsilon(\textbf k)=-2t[\cos(k_x)+\cos(k_y)]-\mu,
\end{eqnarray} 
where $\mu$ is the chemical potential, and $t$ is the nearest-neighbor hopping amplitude. We work with energies in units of the hopping, $t=1$.  We absorb the Hartree shift and restrict our discussion to the $\mu=0$, half-filled case. 

\subsection{Computation Kernel for Second Order Self Energy}
\begin{figure}
    \centering
    \includegraphics[width=0.7\linewidth]{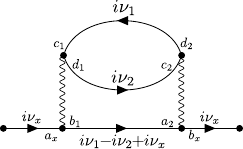}
    \caption{Labelled second order self-energy diagram, $\Sigma^{(2)}$.  The diagram has external frequency, $i\nu_x$, and external band-indices $a_x$ and $b_x$.  Internal band indices are arbitrarily assigned to simplify the form of the interaction matrix elements.}
    \label{fig:sigma}
\end{figure}

We will evaluate the second order self-energy diagram, $\Sigma^{(2)}$, given by
\begin{widetext}
    \begin{equation}\label{eq:o2}
\Sigma^{(2)}=\sum\limits_{i\nu_1,i\nu_2}\sum\limits_{\{a_jb_j\}}G_{a_1b_1}(i\nu_1)G_{a_2b_2}(i\nu_2)G_{a_3b_3}(i\nu_1-i\nu_2+i\nu_x)\bar{U},
\end{equation}
\end{widetext}
where we maintain the subscript notation of Eq.~(\ref{eq:h}).  We will need to interpret the summation over the internal subscripts $a_1b_1$, $a_2b_2$, and $a_3b_3$ for the physical problem being solved and choose a labelling scheme for the diagram. The labelling we will work with is presented in Fig.~\ref{fig:sigma}.   

To generate the computation kernel representation of Eq.~(\ref{eq:primary_final}) we will assign DLR representations to each Green's function in $\Sigma^{(2)}$ in order,
\begin{eqnarray}
    G_{a_1b_1}(i\nu_1)&=\sum_{\ell_1}\frac{A^{\ell_1}_{a_1b_1}}{i\nu_1-x_{\ell_1}}, \\
    G_{a_2b_2}(i\nu_2)&=\sum_{\ell_2}\frac{A^{\ell_2}_{a_2b_2}}{i\nu_2-x_{\ell_2}}, \\
    G_{a_3b_3}(i\nu_1-i\nu_2+i\nu_x)&=\sum_{\ell_3}\frac{A^{\ell_3}_{a_3b_3}}{i\nu_1-i\nu_2+i\nu_x-x_{\ell_3}}.
\end{eqnarray}
After replacing these expressions in Eq.~(\ref{eq:o2}) one can perform the Matsubara summation analytically.  This results in the kernel
\begin{equation}\label{eq:o2_analytic}
    \mathcal{K}^{(2)}(i\nu_x,\{\ell\},\beta)=\frac{[f(x_{\ell_1})-f(x_{\ell_2})][f(-x_{\ell_3})+n(x_{\ell_2}-x_{\ell_1})]}{i\nu_x+x_{\ell_1}-x_{\ell_2}-x_{\ell_3}},
\end{equation}
where $f(x)$ and $n(x)$ are the Fermi and Bose distribution functions respectively. 
We evaluate the kernel, $\mathcal{K}^{(2)}_{i\nu_n,\beta}(\{\ell\})$, for this diagram at a particular inverse temperature $\beta$ for a range of frequencies.  Specifically we obtain the kernel for the first 10 Matsubara frequencies at a dimensionless $\beta=5$.  We place the DLR basis poles and the results of the full kernel in the supplemental files where for each $i\nu_n$ the kernel is a list of complex numbers for a particular choice of basis \cite{supp}.  We have chosen 3 unique DLR representations using $\Lambda=5.15, 5.2,$ and $5.5$ and an error tolerance of $\upvarepsilon\approx 10^{-7}$ at $\beta=5$.  The DLR\cite{kaye:cppdlr} then generates 3 sets of 16 poles, $x_{\ell_1}, x_{\ell_2},$ and $x_{\ell_3}$, which represent the non-physical space used to generate the kernel $\mathcal{K}^{(2)}_{i\nu_n,\beta}(\{\ell\})$.  We generate the kernel using AMI but note that the analytic result of Eq.~(\ref{eq:o2_analytic}) can be used directly\cite{dgaz:2024}.  We emphasize that we are now going to use this one kernel to compute the two distinct physical problems described above.  

In the case of the 2D square lattice Hubbard model (Eq.~(\ref{E:Hubbard})), the non-interacting Green's function is evaluated on an $L\times L$ grid in $k_x$ and $k_y$ and used to generate the DLR coefficients $A^{\ell_j}_{a_jb_j}$.  In this case, $a_j$ and $b_j$ are diagonal states and these indices are reinterpretted to refer to the $k_x$ and $k_y$ coordinates respectively while the interaction amplitudes are given simply by $\bar{U}=U^2$. 
The relation between the subscripts and internal momenta are not unique and we have selected the labelling $a_1,b_1=k_{1x},k_{1y}$, $a_2,b_2=k_{2x},k_{2y}$, and $a_3,b_3=k_{1x}-k_{2x}+k_x,k_{1y}-k_{2y}+k_y$.  Importantly, the constraint on the final indexes is the result of a Fourier transform that leads to momentum conservation but also eliminates and integration from the integration space. This leads to and expression for the coefficients given by
\begin{equation}\label{eq:Chubb}
    C^{(2)}_{\beta,\mathbf{k}}(\{\ell\})=\sum\limits_{\substack{k_{1x},k_{1y}\\k_{2x},k_{2y}}}  A^{\ell_1}_{\mathbf{k_1}}  A^{\ell_2}_{\mathbf{k_2}}  A^{\ell_3}_{\mathbf{k_1}-\mathbf{k_2}+\mathbf{k}} U^2,
\end{equation}
where $\mathbf{k}=(k_x,k_y)$ is the external momentum.

We perform the summation of the internal momenta grids for each $\ell_1,\ell_2$, and $\ell_3$ to obtain the total weights.  If one is interested in isolating the dependence of an observable on the geometry, $L$, one can refine the number of points in the spatial grid.  While this increases the geometry of the subscripts of the $A^{\ell_j}_{a_jb_j}$ coefficients it does not change the geometry of the resulting $C^{(2)}(\{\ell\})$ coefficients, and so the same kernel $\mathcal{K}^{(2)}_{i\nu_n,\beta}(\{\ell\})$ can be used for different system sizes. 

In the case of the multi-band impurity problem (Eq.~(\ref{eq:h})) the coefficients $a_jb_j$ represent diagonal and off-diagonal Green's function matrix element indices. We do a single shot calculation starting from a diagonal Green's function, but write the coefficient without assuming they are diagonal and also include the interaction matrix elements $\bar{U}=U_{a_1b_1c_1d_1}U_{a_2b_2c_2d_2}$ where some indexes must match external indices $a_x,b_x$. 
We assign specific labels to the $a_jb_j$ indices in Fig.~\ref{fig:sigma} and specify which indexes are external lines of the diagram.  This results in
\begin{align}\label{eq:Ch2}
       C^{(2)}_{\beta,a_x,b_x}(\{\ell\})=\sum\limits_{\substack{\{a_1b_1c_1d_1\} \\ \{a_2b_2c_2d_2\}}}  &A^{\ell_1}_{d_2c_1}  A^{\ell_2}_{d_1c_2}  A^{\ell_3}_{b_1a_2}  \nonumber\\ 
       & \times U_{a_1b_1c_1d_1}U_{a_2b_2c_2d_2}\delta_{a_1a_x}\delta_{b_2b_x}.
\end{align}
We will apply this to a two band toy problem, but this expression is valid for any number of bands for a choice of $a_x,b_x$.  We notice that in both Eqns.~(\ref{eq:Chubb}) and (\ref{eq:Ch2}) the coefficients depend on the choice of inverse temperature $\beta$ and also any external spatial degrees of freedom of the system.  As is the case in lattice problems that a variety of distinct $L\times L$ grids could be used while invoking the same Kernel, the choice of the molecular basis also does not impact the use of the same kernel, $\mathcal{K}^{(2)}_{i\nu_n,\beta}(\{\ell\})$. The kernel is truly universal and can be used to generate the solution for the second order self energy diagram for any physical problem.

\begin{figure}
    \centering
    \includegraphics[width=\linewidth]{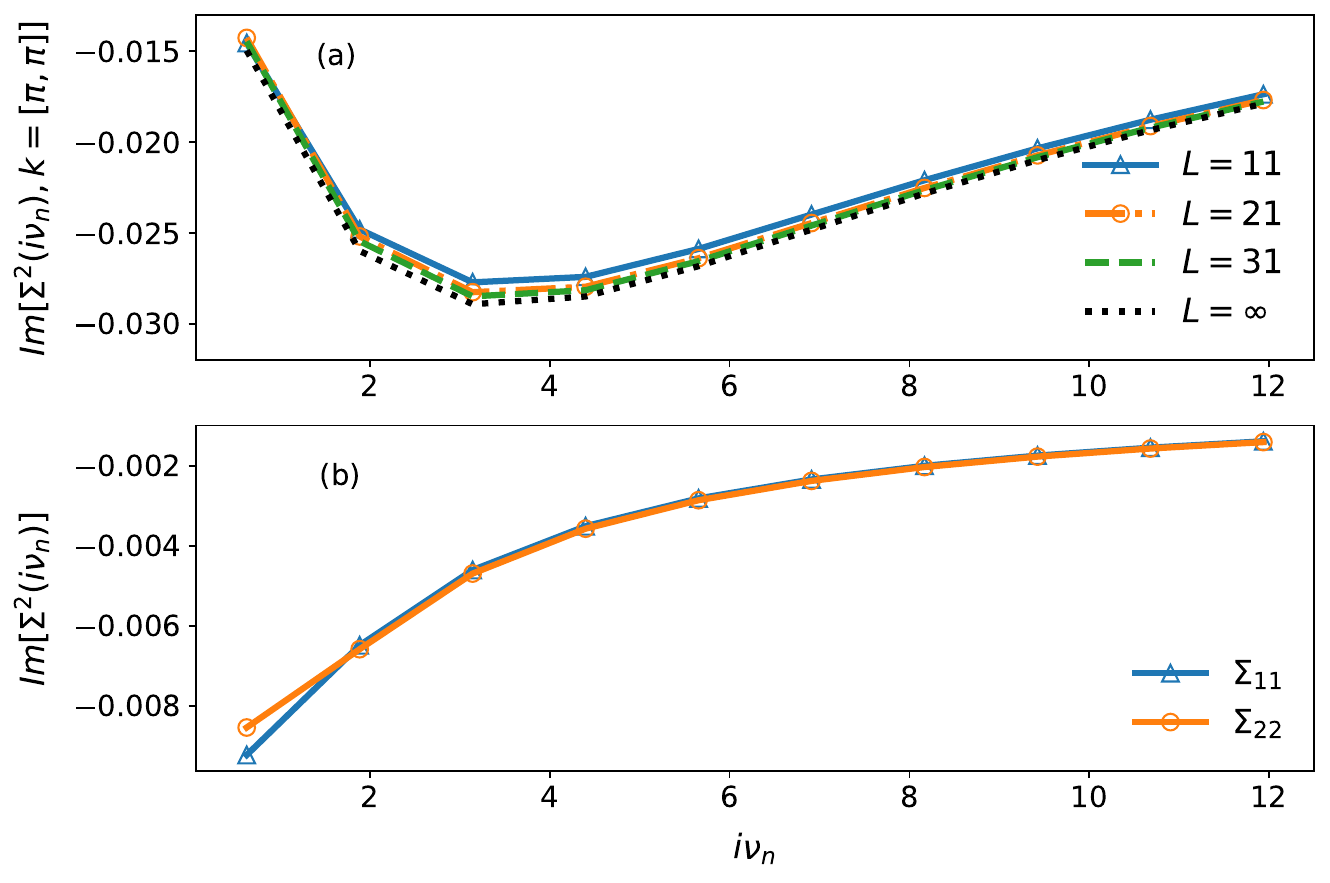}
    \caption{The imaginary part of the second order self energy for the first 10 Matsubara frequencies at $\beta=5$ utilizing a precomputed computation kernel for the diagram. (a) The 2D Hubbard result for external momentum $k_x=(\pi,\pi)$ at $U/t=1$ for the $L\times L$ system with $L=11,21$, and 31.  Also included is a benchmark reference for the $L\to\infty$ limit.   (b) The H$_2$ molecule for the sto-6G basis at equilibrium separation $r=0.74$\AA which exhibits only diagonal elements of the self-energy. }
    \label{fig:resultsdata}
\end{figure}

As a demonstration of the utility of the approach, 
we plot the second order self-energy for our two toy problems in Fig.~\ref{fig:resultsdata}.  The results are obtained via Eq.~(\ref{eq:primary_final}) where the same computation kernel is used for both physically distinct problems.  In the case of the Hubbard model, we choose a single point in the Brillouin zone at an external momentum of $k_x=(\pi,\pi)$.  Since we only need to compute $C^{(2)}_{\beta,k_x, k_y}(\{\ell\})$ it is straightforward to probe increasing system sizes at an expense of $\mathcal{O}(L^{md})$, where $m=2$ is the expansion order and $d$ is the spatial dimensionality.  
Results for $L=11$, 21, and 31 are shown as well as a benchmark for the thermodynamic $L\to\infty$ limit, obtained from continuous stochastic integration methods. 
We emphasize that the results are well behaved and systematically approach the TL benchmark result obtained via continuous stochastic integration.  For each choice of $L$ the approach is a virtually exact numerical evaluation of this diagram for the finite-size problem, where the only uncertainty comes from the DLR fit of the original coefficients and the details of how that uncertainty propagates into the kernel.  In all cases explored, the uncertainty is exponentially suppressed, and since we have selected a DLR representation with an error tolerance of $\upvarepsilon\approx 10^{-7}$ the error in these results is completely negligible on this scale.  Deviation from the $L\to\infty$ result is due only to finite-size effects.  
In Fig.~\ref{fig:resultsdata}(b) we present the diagonal terms of the self-energy for the H$_2$ molecule.  The coefficients $C^{(2)}_{\beta,a_x,b_x}(\{\ell\})$ are obtained via Eq.~(\ref{eq:Ch2}) and then the self energy is generated using the same kernel as for the Hubbard Hamiltonian.   Since the two molecular states $h_{11}$ and $h_{22}$ are not energetically symmetric, there are expected to be small differences between the two curves.  Unlike the lattice problem there are no additional spatial degrees of freedom and so the result is exact for this choice of basis.  One could choose to represent the H$_2$ molecule in a more complex basis, and while this increases the numerical expense of evaluating Eq.~(\ref{eq:Ch2}) the computation kernel is unaffected. 

The problems solved here have little in common.  We are not interested in the specific result for either problem as they can be straightforwardly verified.  Instead, we emphasize that we have a single kernel from which accurate results for the selected diagram can be obtained by only evaluating a trivially parallelizable sum of products of $A^{\ell}_{ab}$ coefficients.  Since this same procedure can be applied to any diagram the potential applications of this approach are endless. 
For transparency, the full details of the kernel $\mathcal{K}^{(2)}_{i\nu_n,\beta}(\{\ell\})$ are provided in the supplementary files along with the problem specific coefficients for each curve in the figure \cite{supp}.

\section{Discussion and Conclusions}
We have applied the discrete Lehmann representation to Feynman diagrams comprised of matrix-valued Green's functions.  This process replaces the physical computational space with an exponentially larger, auxiliary computational space defined by an arbitrary choice of DLR poles.  The byproduct of the projection of the Green's function to an auxiliary space is that the temporal and spatial integrations completely decouple.  The temporal integrals become equivalent to a multi-band impurity problem that can be treated exactly, for a given diagram, via algorithmic Matsubara integration.  The spatial integrals are simplified into simple sums and products of complex numbers that are known, and arise from the DLR fitting procedure.  The key result of this methodology is the ability to precompute the computation kernel a single time and reuse it for alternate physical problems.  
 Once the computation kernel is obtained for a set of external frequencies and temperatures, it can be used to solve that same diagram for any physical problem.  Solving any new physical problem requires only the solution to the spatial components of the diagram that, in the computation kernel approach, will always appear as a product of known tensor elements.  This represents a drastic simplification when solving correlated electron problems - products of complex numbers with no correlation functions nor exponential functions. 

Whether or not there is computational advantage to this approach has yet to be demonstrated.  At minimum, the reuse of the kernel reduced the computational expense of renormalized AMI expansions by the kernel evaluation cost of $r^{2m-1} c^m$.  The kernel is effectively a look-up table for the temporal integrals of Feynman diagrams with a memory cost of $r^{2m-1}$.  For complex doubles at $r=16$ this equates to 64kB of storage at order $m=2$, 16MB at $m=3$ and 4GB at $m=4$.  These values, although not small, are not prohibitive and opportunities for compression have yet to be explored. 
Importantly, the method we present allows for fully self-consistent perturbative approaches that were not previously possible for arbitrary problems, as was already demonstrated for impurity problems with no spatial integrals in Ref.~\cite{dgaz:2024}.  The same self-consistent procedures apply here, where on each step of the self-consistency there is no change to the computation kernel and one simply updates the coefficients $C^{\mathcal{D}}$ to reflect updates to the DLR fit coefficients $A^{\ell_j}_{a_jb_j}$.  Further, diagrams with similar pole structure will likely have identical kernels allowing for potential factorization within each order of the diagrammatic expansion\cite{GIT,kozik:2025:parallel}.

Finally, when one takes a step back from the details of the math there is a simple message in this work.  Abstracting the physical space of poles to a reusable, computational space introduces the possibility of exciting, practical applications.  By having the kernel contain none of the problem specific information, caching or generating a repository of kernels has the potential to reduce computation costs and generate high quality data to a tunable degree of accuracy either as inputs or to validate the predictions of large quantitative models.
Large quantitative models (LQMs) require massive amounts of accurate numerical data and are increasingly used to analyze, predict and simulate complex systems.\cite{alphafold}  The applications of LQMs range from new material discovery for use in batteries and catalysis reactions to genomics research where binding affinities, molecular design and protein-ligand interactions have the potential to revolutionize drug design.  Although more code development is needed in order to apply the computation kernel to this broad spectrum of problems, our approach appears to be a promising avenue for future research.

\section{Acknowledgement}
We acknowledge the support of the Natural Sciences and Engineering Research Council of Canada (NSERC) RGPIN-2022-03882 and (NRC) AQC-200-1. Our codes make use of the ALPSCore\cite{ALPSCore,alpscore_v2} and AMI libraries\cite{libami,torchami}.

\bibliography{refs.bib}

\end{document}